# Scanning mid-IR-laser microscopy: an efficient tool for materials studies in silicon-based photonics and photovoltaics

O. V. Astafiev [*] [†] , V. P. Kalinushkin, and V. A. Yuryev

General Physics Institute of the Russian Academy of Sciences

38 Vavilov Street, Moscow, GSP–1, 117942, Russia

Tel.: +7 (095) 132-8144, Fax: +7 (095) 135-1330, E-mail: vyuryev@kapella.gpi.ru

**Abstract**

A method of scanning mid-IR-laser microscopy has recently been proposed for the investigation of large-scale electrically and recombination-active defects in semiconductors and non-destructive inspection of semiconductor materials and structures in the industries of microelectronics and photovoltaics. The basis for this development was laid with a wide cycle of investigations on low-angle mid-IR-light scattering in semiconductors. The essence of the technical idea was to apply the dark-field method for spatial filtering of the scattered light in the scanning mid-IR-laser microscope together with the local photoexcitation of excess carriers within a small domain in a studied sample, thus forming an artificial source of scattering of the probe IR light for the recombination contrast imaging of defects.

The current paper presents three contrasting examples of application of the above technique for defect visualization in silicon-based materials designed for photovoltaics and photonics which demonstrate that this technique might be an efficient tool for both defect investigation and industrial testing of semiconducting materials.

---

[*] Natural Science Center of GPI of RAS.

[†] Present address: Tokyo University, Department of Basic Science, Komaba 3-8-1, Meguro-ku, Tokyo, 153. E-mail: astf@mujin.c.u-tokyo.ac.jp.

## Introduction.

A number of papers appearing since 1995 [[1]]–[[11]] have introduced a new method of microscopy of semiconductors which has as its basis the same physical approaches that were used in the law-angle mid-IR-light scattering (LALS) technique reported in numerous other publications (see e.g. Refs. [[11],[12]] and papers cited therein). The essence of the technical idea was to apply the dark-field method to the scanning mid-IR-laser microscope to cut away the scattered light components with low spatial frequencies. The cut-off frequency of the spatial filter was optimized to eliminate influence of the probe laser radiation on a detector signal and simultaneously allow large defect images to be transferred through the optical system without drastic suppression of their contrast or intensity [[9],[11]]. If a mid-IR light source of the probe radiation (e.g., $CO_2$, $\lambda = 10.6$ μm, or CO-laser, $\lambda = 5.4$ μm) is used, the microscope enables the imaging of accumulations of free carriers which are usually associated with crystal micron-scale domains enriched with ionized defects and impurities (see Refs. [[11],[12]] and references therein)—we call them large-scale electrically active defect accumulations (LSDAs). As the light scattered with LSDAs forms the images in this basic mode of the instrument, we have designated it scanning LALS or SLALS.

To enable the imaging of crystal regions with enhanced recombination activity (so-called large-scale recombination active defects or LSRDs), the local photoexcitation of the excess carriers within a small domain of the studied sample can be used to form an artificial inhomogeneity in the free carrier distribution which serves as a source of IR light scattering. The intensity of light scattering in this case is proportional to the square of magnitude of the excess-carrier concentration inhomogeneity which in turn is governed by the non-equilibrium carrier lifetime and the surface recombination velocity [[13]]. Thus, the image intensity in this case is completely controlled by the excess carrier local lifetime and local surface recombination velocity in the observation point and the scattering of rays forming the image is induced artificially by the focused pumping laser beam. That is why this recombination contrast imaging mode has been designated optical-beam-induced low-angle light scattering or OLALS. This mode is a direct optical analog of the well-known LBIC (or OBIC) technique but distinct LBIC OLALS requires neither Schottky barrier nor p-n junction formation on a sample.

Two additional sub-modes similar to techniques described in Refs. [[14],[15]] are also available in the microscope when it operates without the dark field. The image is formed in these sub-modes in transmitted rays due to the light absorption by free carriers with (in the optical-beam-induced absorption sub-mode) or without (in the intrinsic absorption sub-mode) sample local photoexcitation.

All the above together with possibilities of the local lifetime determination, LSDA composition analysis, simultaneous OLALS and photoluminescence measurements discussed in Ref. [[11]] and magneto-optical measurements for determination of LSDA conductance type proposed in Ref. [[13]] makes the scanning mid-IR-laser microscope a unique, and relatively inexpensive non-destructive and non-polluting tool for both defect investigation in research labs and industrial material testing and quality inspections.

**Defect images.**

Let us illustrate the above with three contrasting examples.

Figure 1 demonstrates corresponding pairs of SLALS and OLALS micrographs for single-crystalline CZ Si:P wafers ($\rho = 4.5\ \Omega\,cm$) from the CCD manufacturing cycle. The images of LSDAs situated in the bulk of the substrates are seen in all the SLALS pictures as white spots (the whiter the image, the higher the free carrier concentration is). LSRDs located in sub-surface (or sub-interface) layers of the substrates are manifested in the OLALS pictures as dark areas (the darker the image, the lower is the excess-carrier concentration—and also the shorter the lifetime). The latter defects are distributed along the crystallographic directions and seem to be the ones responsible for video defects in CCDs. LSDAs may also give rise to video defects, especially if located in the vicinity of the working layers, although their effect on the video signal of CCDs is doubtful if they are situated deep in the crystal volume.

SLALS and OLALS micrographs of the samples of single-crystalline $Si_{1-x}Ge_x$ alloy with Ge content from 2.2 to 4.7 at. % —a promising material for solar cells—are presented in Fig. 2. Two areas were revealed in the X-ray topographs of these crystals: an area free of striation and dislocations around the wafer centers (area I) and an area containing striation and dislocations in the periphery of the wafers (area II) [[10]]. SLALS pictures shows the striation in area II and no striation in area I. OLALS pictures demonstrate that no or low (Fig. 2 (*h*)) recombination contrast is usually caused by the grown-in striations in the crystal bulk, although a high contrast was revealed in Fig. 2 (*l*). The second type of defects manifested as dark stripes in the OLALS micrographs (Fig. 2 (*b*),(*l*)) can likely be identified as dislocations and dislocation walls which are registered in X-ray patterns of area II and revealed by etching. The last type of defects observed are those seen as black spots in the OLALS patterns (Fig. 2 (*b*),(*d*),(*f*),(*h*),(*j*)). They are present in both areas and have a non-dislocation origin. Some non-dislocation defects were found in both areas by the selective etching which may be similar to those revealed by OLALS. The latter defects seem to be the main lifetime (and cell efficiency) killing extended defects in the studied material.

Scanning mid-IR-laser microscope images of multicrystalline Si for solar cells obtained in the optical-beam-induced absorption sub-mode are depicted in Fig. 3. Dark images of grain boundaries are clearly seen in both pictures. The images are formed due to the lower non-equilibrium carrier concentrations generated by the focused pumping beam in the vicinities of the strongly recombination-active grain boundaries in comparison with those generated when the probe beam scans the sample far from the grain boundaries. As a consequence, a lower concentration of the electron–hole pairs diffuses in the crystal bulk where the probe mid-IR light is absorbed by the free carriers. Figure 3 shows the applicability of the mid-IR-laser microscopy for investigations and inspections of the grain boundary passivation efficiency—well passivated grain boundaries obviously cannot be visualized by means of OLALS or optical-beam-induced absorption.

**Conclusion.**

In conclusion we would like to emphasize once again that scanning mid-IR-laser microscopy might serve as a powerful research and testing tool in scientific and industrial laboratories. Taking advantage of its unique analytical opportunities coupled with non-destructiveness, relatively low cost and ease of use one can investigate or test materials for such specific classes of defects as LSDAs and LSRDs. These have a special importance for such branches of industry as silicon based photonics (e.g., manufacturing of IR photodetector arrays whose yield may obviously be strongly influenced by the presence of both LSDAs and LSRDs in sub-surface regions of wafers) and photovoltaics, as the implementation of silicon based materials with low recombination activity is very important for the development of high efficiency and inexpensive solar cells.

**Acknowledgments.**

The authors express their appreciation to the Ministry of Science and Technologies of the Russian Federation for the financial support of this work which has been carried out within the framework of the sub-program "Perspective Technologies and Devices of Micro- and Nanoelectronics" under the grant No. 02.04.3.2.40.Э.24.

# REFERENCES.

**FIGURE CAPTIONS.**

Fig. 1. SLALS (*a*),(*c*),(*e*),(*g*) and OLALS (*b*),(*d*),(*f*),(*h*) micrographs of the same regions of CZ Si:P wafers from the CCD structure manufacturing cycle (1×1 mm); (*a*),(*b*): initial wafer; (*c*),(*d*): under $SiO_2$ layer (1200 Å thick); (*e*),(*f*): under $SiO_2$ and $Si_3N_4$ layers; (*g*),(*h*): CCD structure.

Fig. 2. Couples of SLALS (*a*),(*c*),(*e*),(*g*),(*i*),(*k*) and OLALS (*b*),(*d*),(*f*),(*h*),(*j*),(*l*) micrographs of the same regions of single crystalline $Si_{1-x}Ge_x$ wafers (1×1 mm); (*a*)–(*d*): p-type CZ Si (100), 4 at. % of Ge; (*e*)–(*h*): p-type CZ Si (111), 4.7 at. % of Ge; (*i*)–(*l*): n-type CZ Si (111), 2.2 at. % of Ge; (*e*),(*f*),(*i*),(*j*) are close to the wafer centers, the rest are far from the centers.

Fig. 3. Scanning mid-IR-laser microscope images of multicrystalline silicon for solar cells (the optical-beam-induced absorption, 4×4 mm). The darker the image, the shorter the lifetime is. Grain boundaries are clearly seen in the pictures.

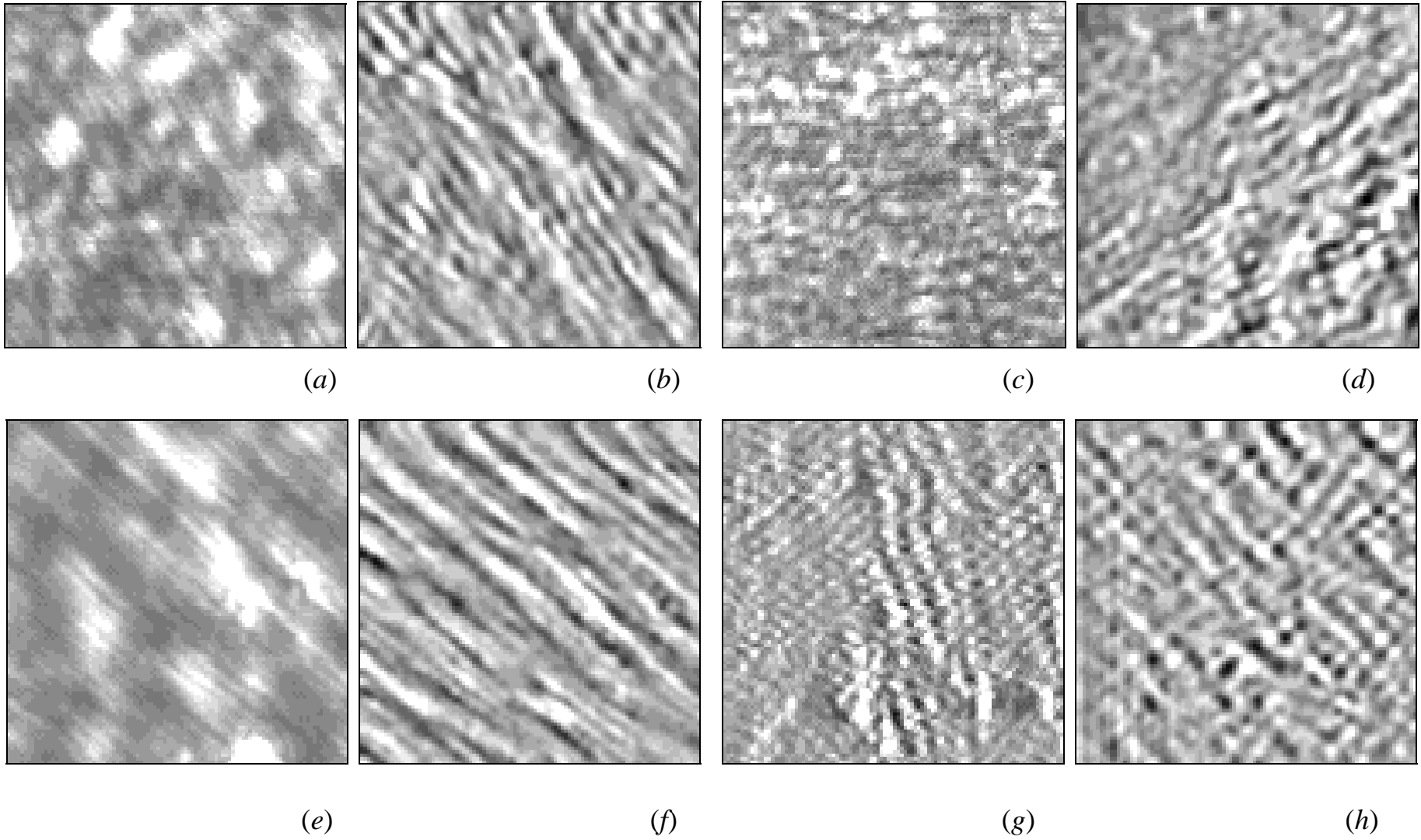

(a)  (b)  (c)  (d)

(e)  (f)  (g)  (h)

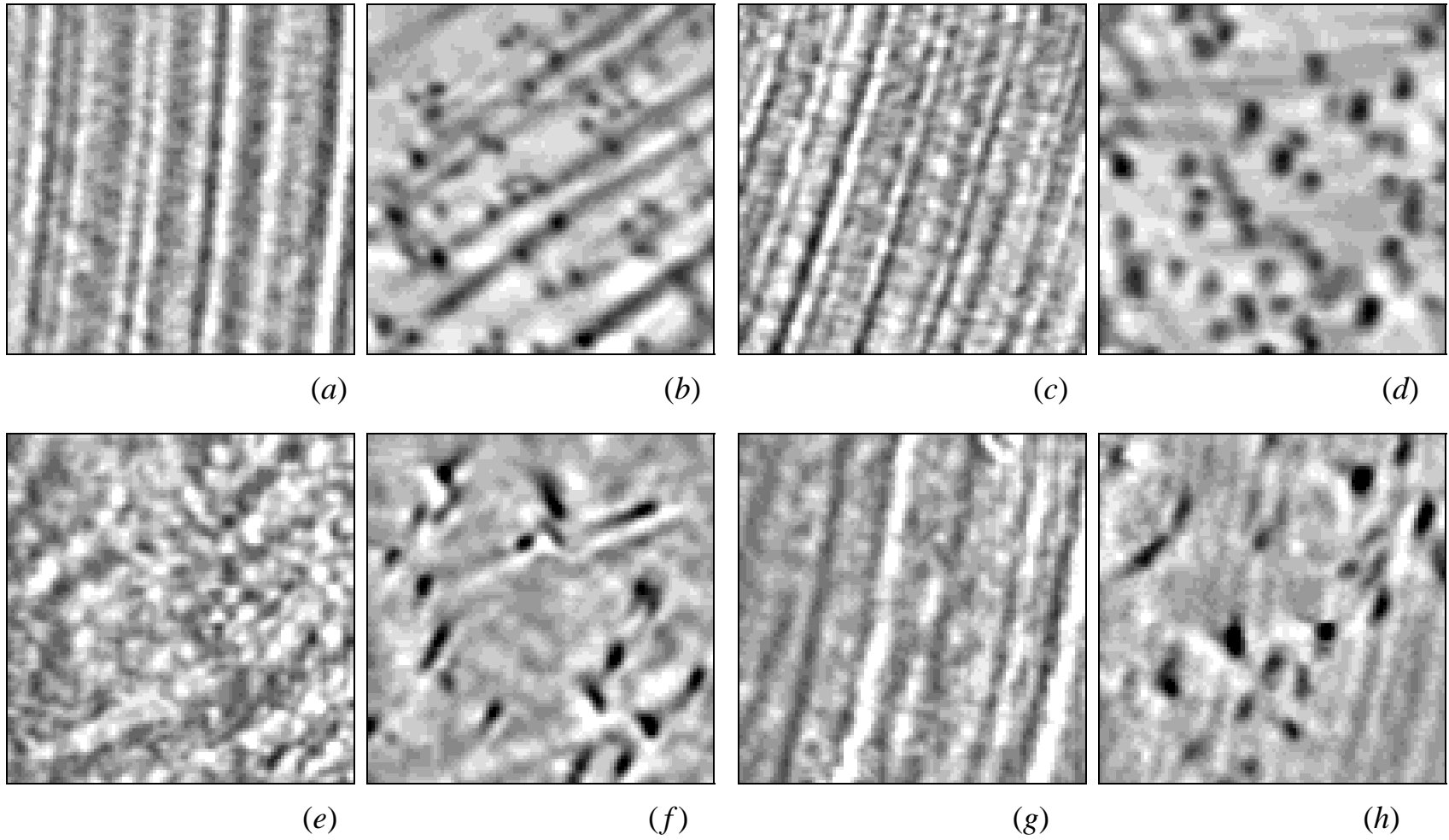

(a)　　　　　　　(b)　　　　　　　(c)　　　　　　　(d)

(e)　　　　　　　(f)　　　　　　　(g)　　　　　　　(h)

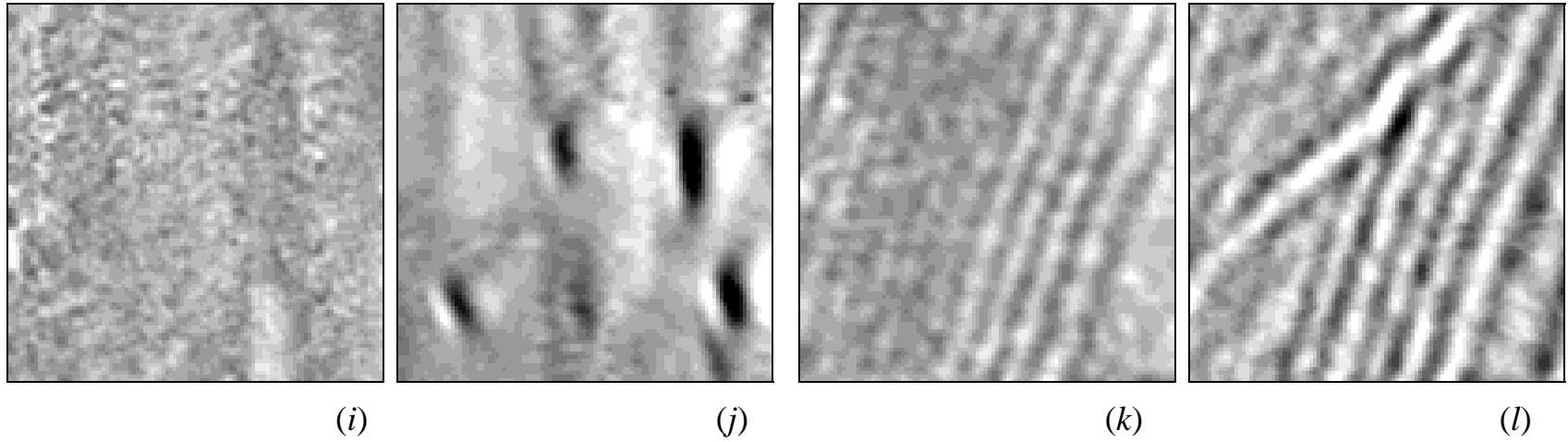

(i) (j) (k) (l)

O. V. Astafiev, V. P. Kalinushkin, V. A. Yuryev.
The Scanning Mid-IR-Laser Microscopy…
Fig. 3.

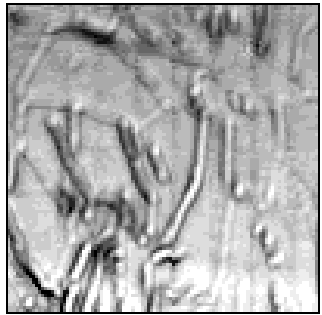 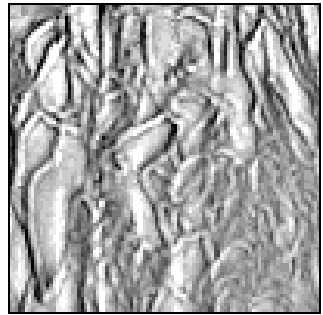

(*a*)  (*b*)